\documentclass[12pt]{article}

\usepackage{epsfig}

\def\Journal#1#2#3#4{{#1} {\bf #2}, #3 (#4)}

\def\NPB{{\em Nucl. Phys.} B}
\def\PLB{{\em Phys. Lett.}  B}
\def\PRL{\em Phys. Rev. Lett.}
\def\PRD{{\em Phys. Rev.} D}

\def\MPLA{{\em Mod. Phys. Lett.} A}

\def\be{\begin{equation}}
\def\ee{\end{equation}}
\def\bea{\begin{eqnarray}}
\def\eea{\end{eqnarray}}

\newcommand{\iai}{I\overline{I}}

\newcommand{\xpr}{{x^\prime}}

\date{}

\begin{document}

\title{{\normalsize\rightline{DESY 96-125}\rightline{hep-ph/9607238}}
    \vskip 1cm
\bf Tracing QCD -- Instantons\\
in Deep Inelastic Scattering\thanks{Talk given
        at the Workshop DIS96 on  
        ``Deep Inelastic Scattering and Related Phenomena'', Rome, Italy,
        April 15-19, 1996; to be published in the proceedings.}\\
\vspace{11mm}}
\author{A. Ringwald and F. Schrempp\\[5mm]
Deutsches Elektronen-Synchrotron DESY, Hamburg, Germany\\[5cm]}

\begin{titlepage} 
\maketitle
\begin{abstract}
We present a brief status report of our broad and systematic study
of QCD-instantons at HERA. 
\end{abstract}
\thispagestyle{empty}
\end{titlepage}
\newpage
\setcounter{page}{2}

\section{Introduction}

Instantons~\cite{bpst} are well known to represent tunnelling 
transitions in non-abelian gauge theories between degenerate 
vacua of different
topology. These transitions induce processes 
which are {\it forbidden} in
perturbation theory, but have to exist in general~\cite{th} 
due to Adler-Bell-Jackiw anomalies.
An experimental discovery of such a novel, non-perturbative manifestation 
of non-abelian gauge theories would clearly be of basic 
significance.

Searches for instanton-induced processes received new impulses
during recent years: First of all, it was shown~\cite{r} 
that the natural exponential suppression of these tunnelling
rates, $\propto \exp (-4\pi /\alpha )$, may be overcome at 
{\it high energies}.
Furthermore, deep inelastic scattering (DIS) at HERA now offers a unique 
window~\cite{bb,rs,grs} 
to experimentally detect processes induced
by {\it QCD-instantons}. Here, a theoretical estimate of the corresponding
production rates appears feasible as 
well~\cite{bb,mrs}, since 
a well defined instanton contribution in the regime of small 
QCD-gauge coupling may be isolated on account of the
photon virtuality $Q^2$. 

In this brief status report we concentrate on a first, preliminary
estimate of the rate for instanton-induced 
events~\cite{mrs} and some characteristics
of the instanton-induced final state along with 
new search strategies~\cite{hws}. 
These new results are based on our instanton Monte-Carlo 
generator~\cite{grs,grs2}
(QCDINS~1.3). 

\begin{figure}
\begin{center}
\epsfig{file=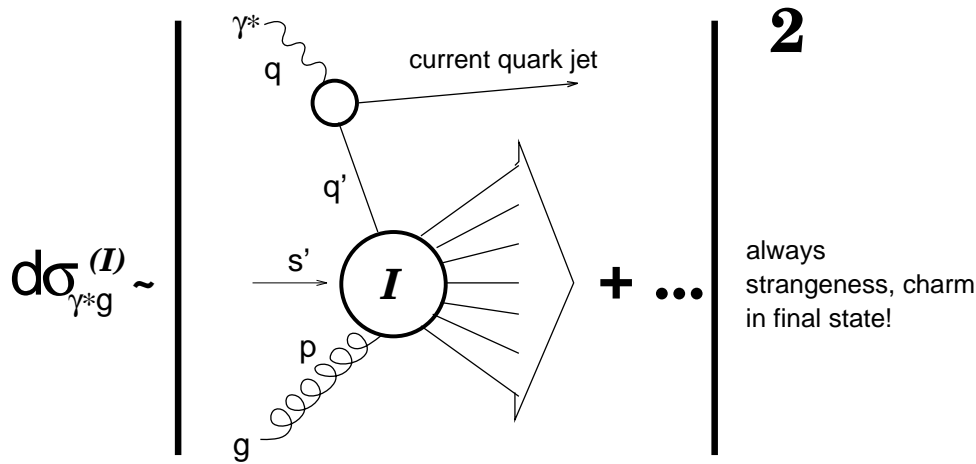,width=6cm,angle=270}
\caption[dum]{\label{f1}
Instanton-induced contribution
to the cross section of $\gamma^\ast g$ scattering.
 }
\end{center}
\end{figure}

\section{Instanton-Induced Cross Sections}

The instanton ($I$) contribution to the nucleon structure functions
is described in terms of the standard convolution of 
parton-structure functions, e.g. ${\cal F}_{2\,{\rm g}}$, 
with corresponding parton densities. The $I$-contribution to the (dominating)
gluon-structure function ${\cal F}_{2\,{\rm g}}$ arises from the 
$\gamma^\ast {\rm g}$ matrix element as displayed in Fig.~\ref{f1}.
The apparent structure of an $I$-subprocess, 
denoted by ``$I$" in Fig.~\ref{f1}, 
is due to the fact that the virtual photon only couples to 
instantons via it's quark content.
We find~\cite{mrs}, that the $I$-contribution to 
the gluon-structure functions may be expressed in terms
of the $I$-subprocess total cross section,
$\sigma_{{\rm q}^\ast {\rm g}}^{(I)}$, 
\begin{equation}
{\cal F}_{2\,{\rm g}}^{(I)} (x ,Q^2) \simeq \sum_{\rm q} e_{\rm q}^2\ x\,
\int_x^1 \frac{dx^\prime}{x^\prime}
\,\left( \frac{3}{8\,\pi^3}\,\frac{x}{x^\prime}\right)\, 
\int_0^{Q^2\frac{x^\prime}{x}}
dQ^{\prime 2} \, 
 \sigma_{{\rm q}^\ast {\rm g}}^{(I)}
(x^\prime, Q^{\prime 2}) \, . 
\label{convolution}
\end{equation}
The integrations in Eq.~\ref{convolution} extend over the
Bjorken variables $Q^{\prime \, 2}=-q^{\prime 2}$ and 
$x^\prime =Q^{\prime 2}/(s^\prime +Q^{\prime 2}) \geq x\geq x_{\rm Bj}$, 
referring to the 
$I$-subprocess.

\subsection{$I$-Subprocess Cross Section}\label{subsec:sigqg}

A standard evaluation~\cite{kr} leads to the following 
result~\cite{mrs},
\begin{equation}
\sigma_{{\rm q}^\ast {\rm g}}^{(I)} (\xpr ,Q^{\prime 2})
\simeq \frac{\Sigma (\xpr )}{Q^{\prime 2}}\ 
\left( \frac{4\pi}{\alpha_s (\mu (Q^\prime ) )}\right)^{21/2}
\ \exp\left[-\frac{4\pi}{\alpha_s (\mu (Q^\prime ) )}
\,F(\xpr )\right]\ 
.
\label{sigma}
\end{equation}
 The running scale $\mu (Q^\prime)$ in
$\alpha_{s}$, satisfying $\mu (Q^\prime)=\kappa\, Q^{\prime } 
\alpha_{s}(Q^\prime)/(4\pi)$ with $\kappa = {\cal{O}} (1)$,
plays the r{\^o}le of an effective renormalization scale.
The $\xpr$ dependence resides in the functions $\Sigma (\xpr)$
and the so-called ``holy-grail" function $F(\xpr)\leq 1$, which are both
known as low-energy expansions in 
$s^{\prime}/Q^{\prime 2}=(1-\xpr)/\xpr\ll 1$ within conventional 
$I$-perturbation theory.
Their form implies a rapid growth of $\sigma_{{\rm q}^\ast {\rm g}}^{(I)}$
for decreasing $\xpr$.  
Unfortunately, in the phenomenologically
most relevant region of small $\xpr$, the perturbative expressions
are of little help and we have to ressort to some
extrapolation. 

A distinguished possibility to go beyond instanton perturbation theory
is the $\iai$-valley 
approximation~\cite{vm1,kr} 
which we have adopted. 
It amounts to the identification of the holy-grail function with
the known $\iai$-valley action.
It appears reasonable to trust this method down to $\xpr= 0.2$,  
where $F(0.2)\equiv S_{\iai-{\rm valley}}(0.2)\simeq 1/2$,  
a value sometimes advocated \cite{hgbound}
as the lower bound for the holy-grail function.
An important phenomenological/experimental task will be to make sure
(e.g. via kinematical cuts to the final state) that $\xpr$ does not
become too small.

Note the following important feature of 
$\sigma_{{\rm q}^\ast {\rm g}}^{(I)}$ as a function of 
$Q^{\prime}$: The $Q^{\prime}$ dependences from the high inverse
power of $\alpha_{s}$ and the exponential in Eq.~\ref{sigma}
compete to produce a strong peak far away from the IR region, 
e.g. $Q^{\prime}_{\rm peak}(\xpr=0.2)\approx 31\,\Lambda$.
This implies that ${\cal F}_{2\,{\rm g}}^{(I)}$, which involves
the integral over $Q^{\prime 2}$ (c.f. Eq.~\ref{convolution}),
is dominated by this peak and hence $Q$ independent (scaling) in the
Bjorken limit. The predicted approach to this scaling limit
resembles a ``fractional twist" term, where the twist is sliding
with $x$: the scaling violations vanish faster for increasing
$x$. 

\subsection{HERA Cross Section}

\begin{figure} 
\vspace{-0.4cm}
\begin{center}
\epsfig{file=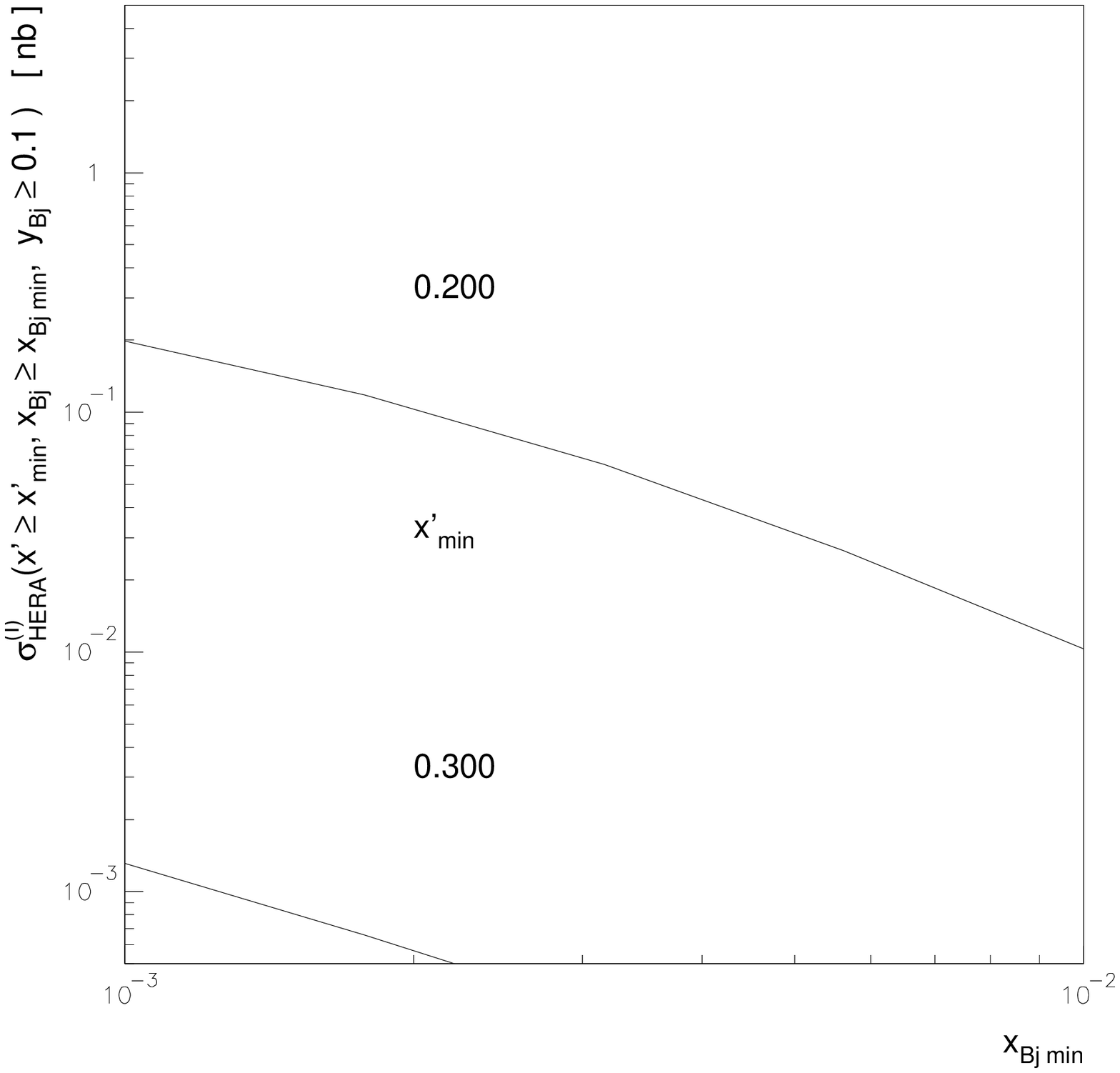,bbllx=17pt,%
bblly=149pt,bburx=536pt,bbury=647pt,height=6.5cm}\hfill
\epsfig{file=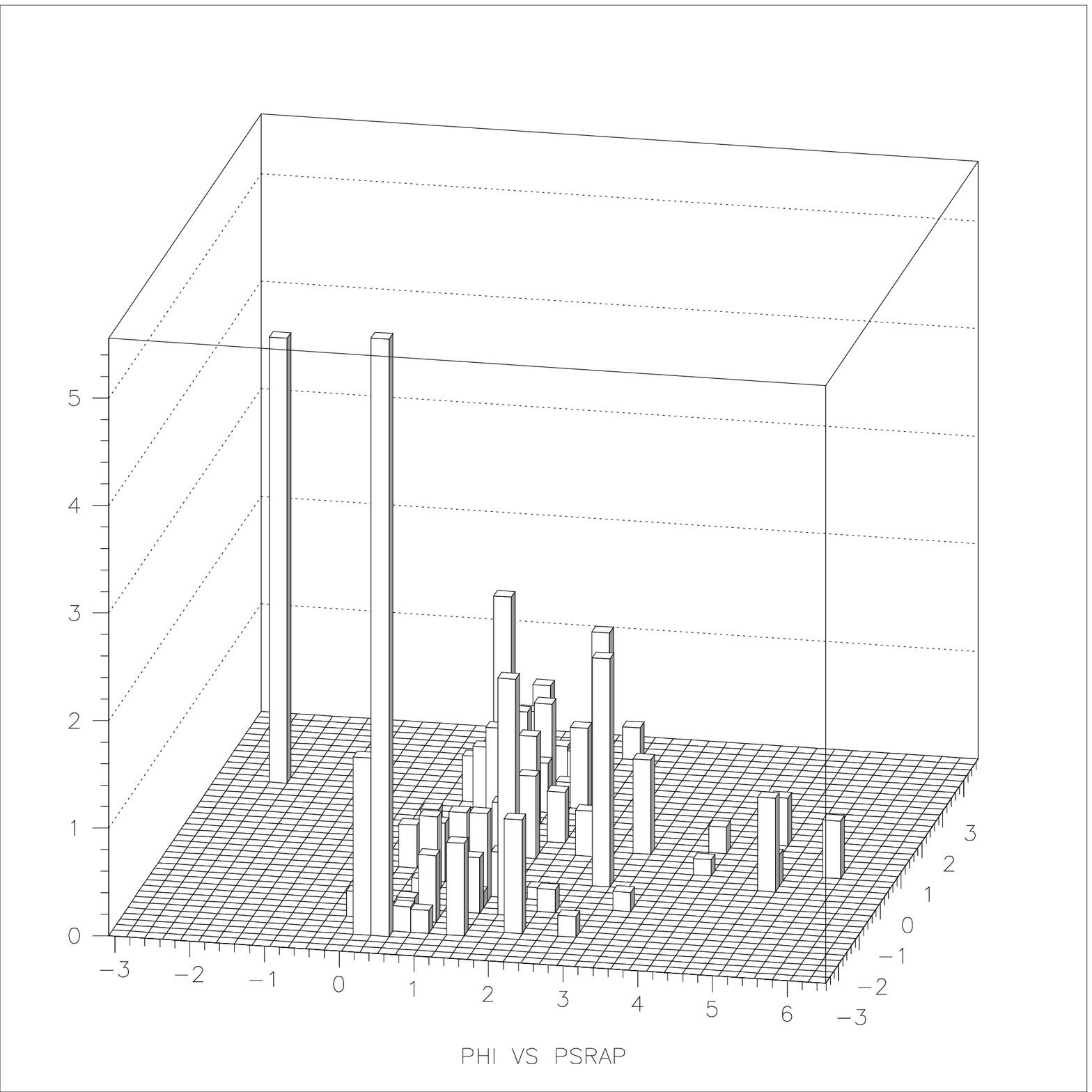,height=6.5cm}
\caption[dum]{\label{f2}
Left: $I$-induced total cross section for HERA (preliminary) with various
cuts. 
Right: Lego plot $(\eta_{\rm lab},\phi_{\rm lab},E_{T}[{\rm GeV}])$ 
of a typical $I$-induced event in the HERA-lab system at $x_{\rm Bj}=~10^{-3}$.
 }
\end{center}
\end{figure}

In Fig.~\ref{f2} (left) we present the resulting $I$-induced total cross
section for HERA for two values (0.2,0.3) of the lower $\xpr$ cut  
(c.f. discussion in Sec.~\ref{subsec:sigqg}), as a function of the
minimal Bjorken $x$, $x_{\rm Bj\ min}$, considered. It is surprisingly
large. 
So far, only the (dominating) gluon contribution has been taken into account. 
The inherent uncertainties associated with the renormalization/factorization
scale dependences may be considerable and are presently being investigated.
Therefore, Fig.~\ref{f2} is still to be considered preliminary.

\section{Final-State Signatures and New Search Strategies}

The typical event (Fig.~\ref{f2} (right)) 
from our Monte-Carlo generator~\cite{grs2} QCDINS~1.3 based on HERWIG~5.8,
illustrates most of the important features characteristic for the
underlying instanton mechanism:
A current-quark jet along with a densely populated hadronic
``band" of width $\triangle \eta =\pm 0.9$ in the 
$(\eta_{\rm lab},\phi_{\rm lab})$-plane~\cite{rs}. 
The band reflects the isotropy
in the $I$-rest system. The total $E_T={\cal O}(20)$ GeV is large as
well as the multiplicity, $N_{\rm band}={\cal O}(25)$. Finally, 
there is a characteristic flavor flow: All (light) flavors are
democratically represented~\cite{th} in the final state. Therefore,
strongly enhanced rates of $K^{0}$'s and $\mu $'s 
(from strange and charm decays) represent crucial signatures 
for $I$-induced events.

A first, preliminary 95\% CL upper limit 
of 0.9 nb on the $I$-induced cross section at HERA 
has been obtained by the H1 collaboration by searching for an excess
in the $K^{0}$ rate~\cite{h1}.

\begin{figure}
\vspace{-0.4cm}
\begin{center}
\epsfig{file=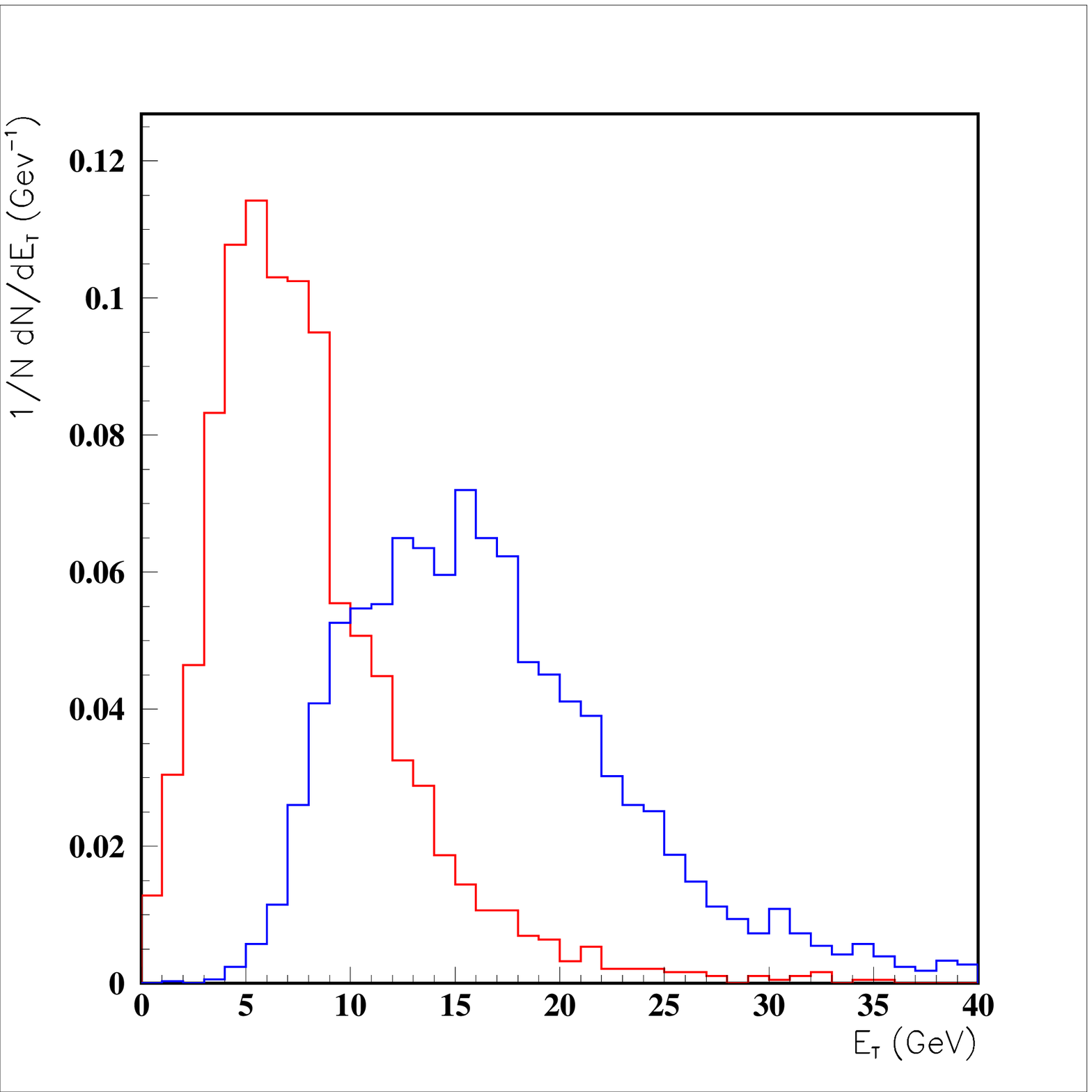,width=5cm,height=6.5cm}\hfill
\epsfig{file=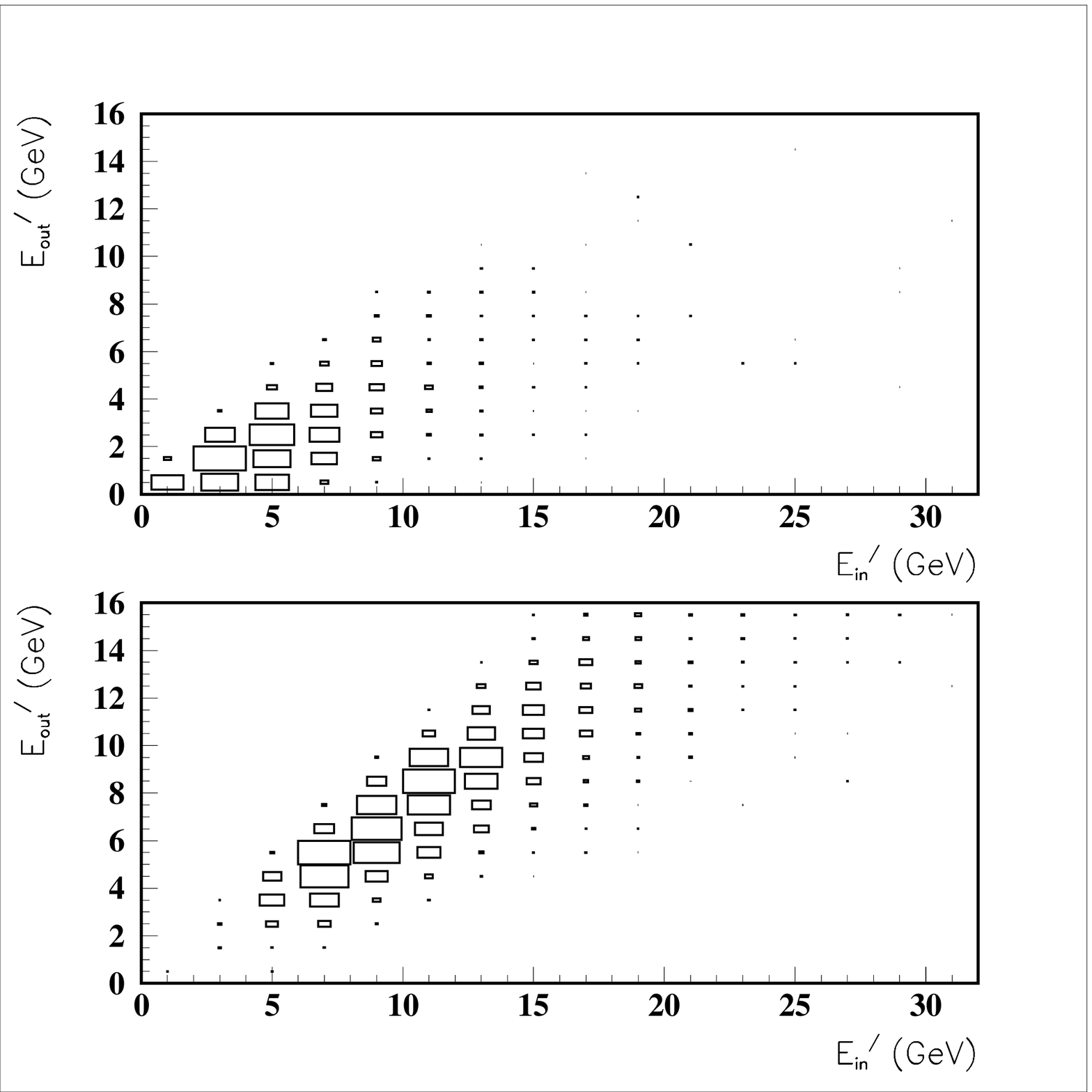,width=8.5cm,height=6.5cm}
\caption[dum]{\label{f3}
Left: $E_{T}$ distributions in the $\gamma^{\ast}$-proton c.m. system for 
normal DIS (left) and $I$-induced events (right). 
Right: $E_{\rm out}$ vs. $E_{\rm in}$ distributions in the 
$\gamma^{\ast}$-proton c.m. system for normal DIS events (top) 
and $I$-induced events (bottom). The primes indicate additional cuts in
$\eta$ to minimize NLO perturbative QCD effects.  
 }
\end{center}
\end{figure}

Let us finally mention some recent attempts~\cite{hws} 
to improve the sensitivity to $I$-induced events by adding in 
characteristic information on the {\it event shapes}.
The first step consists in boosting to the $\gamma^{\ast}$-proton c.m. 
system and looking for events with high $E_{T}$ 
(c.f. Fig.~\ref{f3} (left)). 
We note that in this system 1 and 2 jet (hard) perturbative
QCD processes deposit their energy predominantly in a {\it plane} passing
through the $\gamma^{\ast}$-proton direction. 
In contrast, the energies from $I$-induced events are always distributed 
much more {\it spherically} (isotropy in the $I$-rest system!).
Therefore, one may substantially reduce the normal DIS background 
by looking at
\begin{equation}
E_{\rm out}={\rm min}\,\sum_{i}^{N} \vec{E}_{i}\cdot \vec{n} \, ,
\end{equation}
i.e. by minimizing $E_{\rm out}$ by choice of $\vec{n}$, normal to the
$\gamma^{\ast}$-proton direction. For standard boson-gluon fusion
2 jet events, $E_{\rm out}$ is given by the jet widths. In contrast, for
$I$-induced events $E_{\rm out}\simeq\sqrt{s^{\prime}}/2$ is large.
The quantitative results from the Monte-Carlo simulation, subject to additional
cuts in $\eta$ which are to minimize higher-order perturbative QCD effects,
are displayed in Fig.~\ref{f3} (right). They fully confirm the qualitative
expectations.


\end{document}